# Stable high-pressure phases in the H-S system determined by chemically reacting hydrogen and sulfur

Alexander F. Goncharov[1,2], Sergey S. Lobanov[2,3], Vitali B. Prakapenka[4], Eran Greenberg[4]

[1]Key Laboratory of Materials Physics, Institute of Solid State Physics CAS, Hefei 230031, China

[2]Geophysical Laboratory, Carnegie Institution of Washington, Washington, DC 20015, USA

[3]Sobolev Institute of Geology and Mineralogy, Siberian Branch Russian Academy of Sciences, 3 Pr. Ac. Koptyga, Novosibirsk 630090, Russia

[4]Center for Advanced Radiations Sources, University of Chicago, Chicago, IL 60632, USA

**Synchrotron X-ray diffraction and Raman spectroscopy have been used to study chemical reactions of molecular hydrogen ($H_2$) with sulfur (S) at high pressures. We find theoretically predicted *Cccm* and *Im-3m* $H_3S$ to be the reaction products at 50 and 140 GPa, respectively. *Im-3m* $H_3S$ is a stable crystalline phase above 140 GPa and it transforms to *R3m* $H_3S$ on pressure release below 140 GPa. The latter phase is (meta)stable down to at least 70 GPa where it transforms to *Cccm* $H_3S$ upon annealing (T<1300 K) to overcome the kinetic hindrance. *Cccm* $H_3S$ has an extended structure with symmetric hydrogen bonds at 50 GPa and upon decompression it experiences a transformation to a molecular mixed $H_2S$-$H_2$ structure below 40 GPa without any apparent change in the crystal symmetry.**

Superconductivity at temperature close to 200 K in the H-S system at high pressures (P) has been reported recently [1, 2]. This discovery attracted a great deal of excitement and stimulated a number of experimental and theoretical studies. The key question is clearly about the mechanism of superconductivity, where a conventional electron-phonon coupling scenario [2-6] is challenged by more exotic (complex) theories [7-9]. However, the experimental reports on the relevant to superconductivity properties remain scarce [1, 2, 10-12] making it difficult to clarify this discussion.

Notably, the high-pressure chemical structure and stable composition of materials in the H-S system remain under debate. Indeed, while many theoretical studies consider *Im-3m* $H_3S$ as the superconducting phase near 150 GPa, there are other reports that challenge the decomposition scenario of $H_2S$ [13] or suggest a compound with a modified crystal chemistry SH-$H_3S$ [14]. X-ray diffraction (XRD) experiments on $H_2S$ starting material to 140 GPa at 295 K reported a mixture of phases, where *R3m* (a lower P modification of *Im-3m*) $H_3S$ was a minor phase, but it could not be identified uniquely due the absence of characteristic XRD peaks [15]. Likewise, XRD experiments on $H_2S$ compressed at about 180 K [11, 12]) confirmed the stability of *Im-3m* $H_3S$ but in a phase mixture, where the presence of H-S compounds with other compositions could be substantial likely affecting the superconducting transition temperature ($T_C$)[16]. Most recently, experiments which employed direct elemental synthesis [17] reported only the *Cccm* $H_3S$ phase up to 160 GPa, thus challenging the theory-based view of the superconducting phase (*Im-3m*) and suggested it to be metastable.

Experiments aiming to uncover the properties of the H-S system related to superconductivity have to be performed at high pressures approaching 150 GPa which is challenging because the sample dimensions are necessarily small (<50 μm linear and <10 μm thick). In addition to this, the sample prepared following recipes of Ref. [2] are multiphase in nature [11, 12], making it difficult to draw definitive conclusions about the structure, vibrational properties, conductivity, optical and other properties. Here we present the synthesis of $H_3S$ materials performed out of elemental reagents – sulfur and molecular hydrogen (see also Ref. [17]). We show that the first principles theoretical studies of structure and composition [3, 12, 15, 18-20] correctly predict *Im-3m (R3m)* $H_3S$ compound as the most stable at 110-150 GPa (cf Ref. [17]); thus, our works resolves the controversy about the superconducting phase in the H-S system.

We performed the experiments in laser heated diamond anvil cells (DAC) with 50 to 300 μm central culets. Small pieces of single crystal S have been positioned in a hole in a rhenium gasket and filled with $H_2$ gas at ~ 150 MPa. Subsequently, loaded DACs were cooled down to approximately 180 K, brought to a desired pressure (approximately 50 and 150 GPa), and laser heated, while the anvils were kept at the low temperature near 180 K [12, 21]. This P-T path has been chosen to avoid diamond failure due to the penetration of hydrogen into diamond anvils, as is often the case in DACs experiments on hydrogen. Successive laser heatings of the synthesized $H_3S$ compounds have been also performed at room temperature when needed. Synchrotron XRD measurements at GSECARS, APS, ANL [22] with the X-ray beam spot size as small as 3 x4 μm was used to probe physical and chemical states of the sample. Vibrational properties of $H_3S$ and chemical composition of the samples were probed on decompression off line (at HPSynC and Geophysical Laboratory). Pressure was measured using the Au XRD standard [23] (where available), Raman of the stressed diamond [24], and frequency of the $H_2$ vibron [25]. Two experiments up to 150 GPa showed identical results: cubic bcc-like *Im-3m* phase is synthesized after laser heating (<1300 K) at 190 and 295 K. Small amounts of *R3m* S were preserved in the cavity. Fig. 1(a) presents XRD results at the sample position where contributions from sulfur and rhenium hydride (formed at the gasket/sample interface) were minimized. The formation of a bcc S lattice in $H_3S$ is clearly seen with the lattice parameter (*a*=3.100(5) Å at 140(5) GPa), which is very close to the theoretically calculated value of *a*=3.088 Å for *Im-3m* $H_3S$ (*e.g*. Ref. [3]). The Raman spectra of this phase did not yield any measurable signal, in agreement with the Raman selection rules for *Im-3m* $H_3S$ predicting zero Raman active modes. Please note that the diffraction lines of *Im-3m* $H_3S$ synthesized from S and H are very narrow, much sharper than those observed before in the experiments where $H_2S$ was the initial reactant [11, 12] (Fig. 1(b)). Given the perfect agreement between the positions of the observed and predicted Bragg peaks, we conclude that *Im-3m* $H_3S$ was synthesized following both reaction pathways, but it is much better crystallized in the case of the direct synthesis from the elemental components. We tentatively explain this by a very large diffusivity of hydrogen, which was in excess in this process. In contrast, synthesis of $H_3S$ out of $H_2S$ occurs at the conditions of hydrogen deficiency, which may suppress the formation of a high-quality sample even with laser annealing. The explanation of these broadening of the diffraction lines has been given in terms of the Magnéli" phases – interpenetrating metastable phases of variable composition on a microscopic scale [26]. Unlike previous studies, here we demonstrate the synthesis of the perfectly crystalized *Im-3m* $H_3S$.

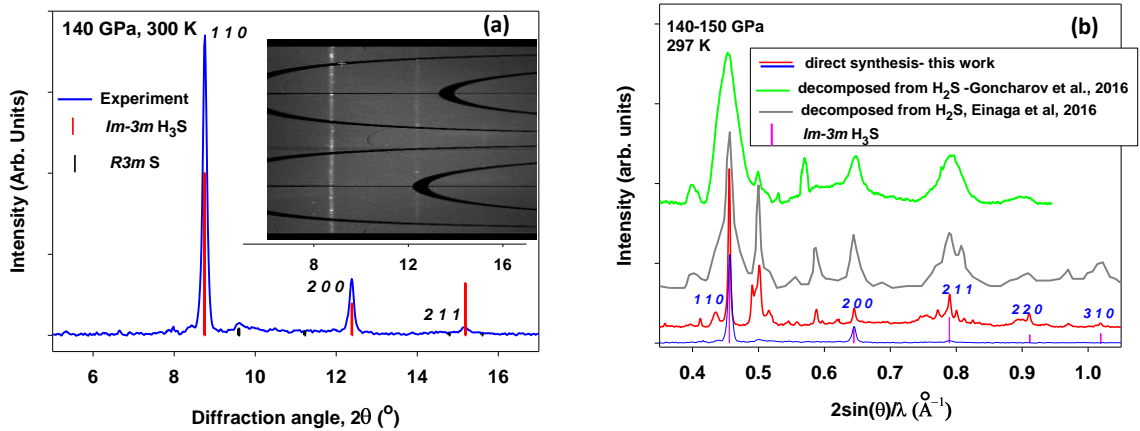

**Fig. 1(a). XRD patterns of *Im-3m* $H_3S$ directly synthesized from $H_2$ and S. The X-ray wavelength is 0.3344 Å. The inset is the raw diffraction images in rectangular coordinates (cake); (b). Comparison of the diffraction patterns of $H_3S$-rich materials prepared by different techniques: direct synthesis (two different experiments are shown in red and blue)- this work and from $H_2S$ precursor [11, 12]. The patterns are plotted as a function of $2 \sin\theta / \lambda$, where $\theta$ is the diffraction angle and $\lambda$ is the X-ray wavelength.**

Unloading of the newly synthesized *Im-3m* H₃S was performed at room temperature monitoring both XRD and Raman spectra. We find that the Bragg peaks split on unloading down to 120 and 138 GPa in two different experiments. The diffraction peak positions match well *R3m* H₃S (Fig. 2(a)), however the distortion of this rhombohedral structure is much larger than predicted theoretically [3] (Fig. 2(b)). The splitting of the (110) has been also observed in the previous investigation of H₃S prepared from H₂S [11] (albeit much less clearly), where *Im-3m (R3m)* H₃S was observed to be metastable down to 92 GPa. The transformation to the *R3m* structure on unloading was proposed based on the decline of the $T_C$, but the authors of Ref. [11] found no evidence of the phase transition in their XRD patterns. Our XRD measurements on directly synthesized *Im-3m* H₃S show a rhombohedral distortion $r=\sqrt{8/3}c/a-1$ increasing first with the pressure decrease and then leveling at the 0.03 value below 110 GPa (Fig. 2(b)). Please note that the diffraction peaks measured in this work are much narrower than in the previous study [11], so we believe that our observations of *r* of is not in disagreement. These results challenge the theory predicting a very small rhombohedral distortion. Alternatively, the observed splitting may be due to nonhydrostatic pressure conditions upon the pressure release. However, we note that two experiments with a very different hydrogen content in the DAC chamber yielded very similar results, suggesting that these observations are characteristic of *R3m* H₃S. No Raman activity has been detected in this phase, although one would expect many Raman active modes due to the asymmetry of the hydrogen bonds in *R3m* H₃S [27]. However, *R3m* H₃S is predicted to be metallic [3, 11] (we also observed sample strongly reflecting visible light at these conditions, i.e., it appears shiny), which makes these modes weak and thus difficult to detect.

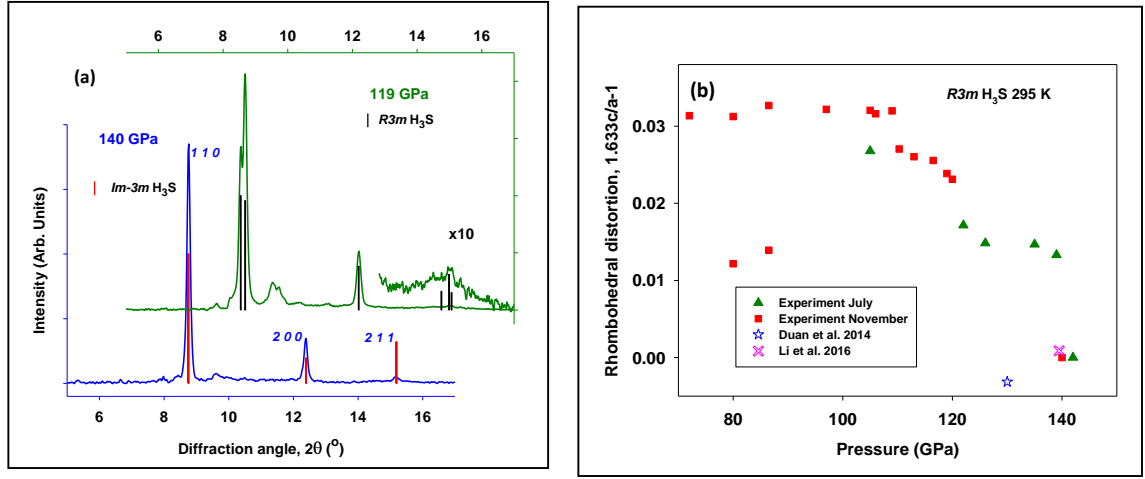

**Fig. 2(a). XRD patterns of *Im*-3*m* H$_3$S directly synthesized from H$_2$ and S and of *R*3*m* H$_3$S measured on pressure release of the same sample. The X-ray wavelength is 0.3344 Å. Please not different horizontal axes for the top and the bottom traces; (b). The rhombohedral distortion of *R*3*m* H$_3$S measured in two experiments $r=\sqrt{8/3}c/a$-1, where $a$ and $c$ are the lattice parameters in the hexagonal setting. A variety of distortions have been observed below 90 GPa depending on the observation points.**

Below 100 GPa or so we find a weakening and broadening of the diffraction peaks assigned to *R*3*m* H$_3$S and also large variations in the value of the splitting depending on the probed sample area (Figs. 2(b), 3(a)). The sample densities (Fig. 3(b)) show a large scattering in values and deviate from the monotonous decompression trend. At approximately 70 GPa we again laser heated the sample below 1300 K and found it transforming instantaneously to another phase, which was clearly identified as *Cccm* H$_3$S, previously predicted theoretically [3] and observed in several different experimental arrangements [12, 17, 28]. These include its formation at low pressures from the gas phases [28], formation by laser heating of H$_2$S at 55 and 110 GPa [14], and synthesis from S and H$_2$ at 75 to 100 GPa [17]. Here we found yet another path -phase transformation of *R*3*m* H$_3$S on unloading. Previously we have reported on the *Cccm* to *R*3*m* transition in H$_3$S (rather sluggish) [12] on pressure increase at 295 K that occurs above 110 GPa covering a large phase coexistence range up to 150 GPa.

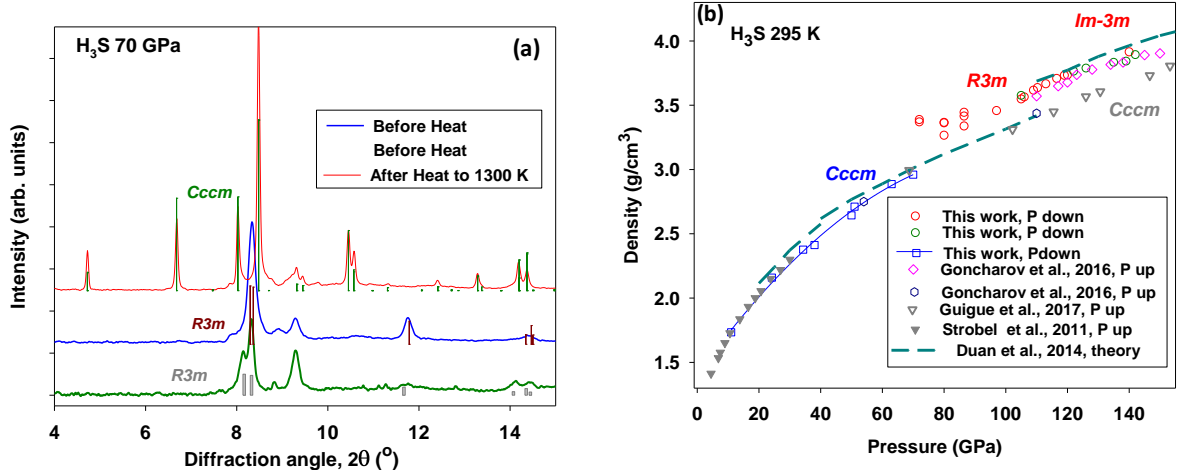

**Fig. 3(a). XRD patterns of *R3m* H$_3$S at 70 GPa before laser heating in two sample positions in comparison to the pattern of *Cccm* H$_3$S synthesized after the heating. The X-ray wavelength is 0.3344 Å; (b). The density of various structural modifications of H$_3$S as a function of pressure measured in the experiments of this work in comparison to those reported in other experiments [12, 17, 28] and calculated theoretically [3].**

Three experiments were performed to investigate the pressure dependent properties of *Cccm* H$_3$S synthesized directly from H$_2$ and S. In one of these runs, we have continued unloading the sample traversing the *R3m-Cccm* transition as described above following decompression-induced changes in Raman spectra. In the other two, we documented a direct synthesis of *Cccm* H$_3$S at approximately 50 and 60 GPa independently complementing observations of Ref. [17] at higher pressures (Fig. 3(b)). In one of these two experiments we managed to concomitantly measure XRD and Raman spectra on unloading. XRD measurements show that the *Cccm* H$_3$S structure adequately describes the data down to 10 GPa, and we find no irregularities in the equation of state (EOS) (Fig. 3(b)). However, we found substantial changes in the Raman spectra at 17 and 40 GPa (Fig. 4), at which molecular ordering and/or orientations modify. At 17 GPa, in parallel to the previous observations of Ref. [28], the S-H and H-H stretching modes split suggesting orientational ordering of $H_2S$ and $H_2$ molecular units in in the structural modification *Cccm*-II. In addition to this, above 40 GPa we find a drastic decrease in intensity and a reduction of a number of the observed S-H and H-H stretching modes down to one of each (Fig. 4). Concomitantly, we observed a visual change in the sample appearance: it becomes shiny in the reflected light. It is interesting that the low-frequency spectra of all three modifications of *Cccm* $H_3S$ are similar yielding just one weakly pressure dependent band at approximately 500 cm$^{-1}$. At above 40 GPa this mode is dominates in the Raman spectra in agreement with our previous report [12], where we synthesized *Cccm* $H_3S$ from $H_2S$.

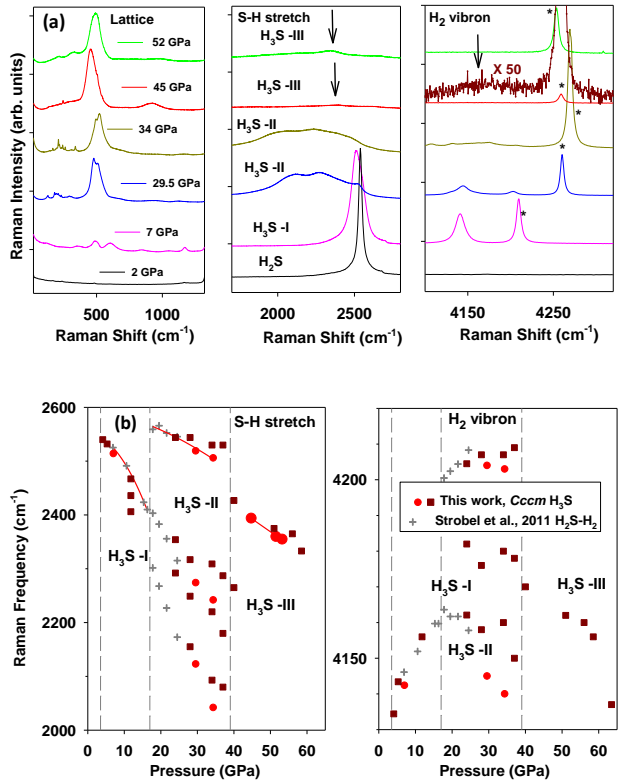

**Fig. 4(a). Raman spectra of *Cccm* $H_3S$ measured on pressure unloading. The excitation wavelength is 532 nm. The asterisks mark the vibron of pure $H_2$; the arrows mark weak peaks; (b). The**

pressure dependence of Raman frequencies of *Cccm* $H_3S$ as a function of pressure. The results of this work are compared to those of Ref. [28].

Our experiments provide strong evidence for pressure-induced destabilization of $H_2S$ in favor of $H_3S$ as well as the theoretically predicted sequence of phase transitions *Cccm - R3m - Im-3m* $H_3S$ [3], in order of increasing pressure. We found that *Cccm* $H_3S$ is stable in a wide pressure range (see also Ref. [17]), but unlike the observations of Ref. [17] we found that *Im-3m* $H_3S$ is stable in the limit of high pressures (140 GPa). This has been demonstrated in two different experimental arrangements: in the presented here direct synthesis of *Im-3m* $H_3S$ from $H_2$ and S (two experiments), and also in the *Cccm* to *Im-3m* transformation observed at 295 K in the experiments where $H_2S$ was a reactant [12]. The discrepancy in the phase relation is likely due to a very large stability/metastability range of *Cccm* $H_3S$. Unlike the work reported in Ref. [17], we have synthesized *Im-3m* $H_3S$ directly from $H_2$ and S at 140 GPa. In the experiments [11, 12], low crystallinity *Im-3m* $H_3S$ was synthesized from a heterogeneous disperse phase mixture (which included *Cccm* $H_3S$) for which it was likely easier to overcome the kinetic barrier of transformation.

As predicted theoretically [3, 27], superconducting *Im-3m* $H_3S$, which has symmetric hydrogen bonds, breaks the symmetry at lower pressures yielding *R3m* $H_3S$, but the distortion is quite small. On the contrary, our experiments demonstrated a quite large distortion, mechanism of which is unclear. It is likely that this distortion along with the symmetry breaking contributes into the superconductivity decline reported in Ref. [11].

Our observations of an additional modification of *Cccm* $H_3S$ (III) above 40 GPa is likely related to a possible band gap closure of this material under pressure suggested to occur at 110 GPa [3]. Our Raman results (Fig. 4) suggest that this transformation must be also coupled to a subtle structural modification, where two crystallographically different $H_2$ molecules become identical or indistinguishable vibrationally. All three *Cccm* modifications show lower $H_2$ vibron frequencies compared to that of the pure molecular $H_2$ (and very broadened band in *Cccm*–III (Fig. 4(a)) suggesting that an association with the $H_2S$ molecules [28] persists to high pressures (up to 60 GPa). It is interesting that the high-pressure modification *Cccm*-III $H_3S$ has a single S-H stretch mode. Based on all these observations, we speculate that formation of *Cccm*-III is associated with the hydrogen bond symmetrization within the $H_2S$ subsystem (*e.g.* Ref. [29]) transforming the molecular host-guest $H_2S$-$H_2$ to an extended structure. The low-frequency mode at 500 cm$^{-1}$, which shows a slight red shift in *Cccm*-III, represents vibrations of tetrahedrally coordinated S.

In conclusions, we have synthesized all three theoretically predicted high-pressure modifications of $H_3S$ at pressures that correlate well with the superconducting properties. Unlike previously reported synthesis of $H_3S$ by decomposition of $H_2S$, a direct synthesis from $H_2$ and S results in the formation of pure $H_3S$. We suggest that superconducting properties of the directly synthesized $H_3S$ should be examined as such prepared material has a much better crystallinity compared to that obtained by $H_2S$ decomposition.

A.F.G. was partly supported by the Chinese Academy of Sciences visiting professorship for senior international scientists (Grant No. 2011T2J20) and Recruitment Program of Foreign Experts. A.F.G. and V.P. are grateful to the NSF MRI EAR/IF1531583 award. S.L. was partly supported by state assignment project No. 0330- 2014-0013. GSECARS is supported by the U.S. NSF (EAR- 0622171, DMR-1231586) and DOE Geosciences (DE-FG02-94ER14466). Use of the APS was supported by the DOE-BES under Contract No. DE-AC02-06CH11357. We thank